\RequirePackage[colorlinks = true, urlcolor   = blue, citecolor  = blue]{hyperref}
\documentclass[draft]{agujournal2019}
\usepackage{url} 
\usepackage{lineno}
\usepackage[inline]{trackchanges} 
\usepackage{soul}
\draftfalse

\graphicspath{{figures/}}
\usepackage{amsmath}
\usepackage[inkscapeformat=png]{svg}
\usepackage{calrsfs}

\DeclareMathOperator{\ra}{\mathit{Ra}}

\usepackage{color}
\definecolor{ForestGreen}{RGB}{34,139,34}
\usepackage{graphicx}
\usepackage{tikz}
\usetikzlibrary{positioning} 
\usepackage{xr}
\journalname{Geophysical Research Letters}

\begin{document}
\title{How modeling assumptions shape predictions of convective mixing of carbon dioxide}
\authors{Marco De Paoli\affil{1}  and Sergio Pirozzoli\affil{2}}

\affiliation{1}{Institute of Fluid Mechanics and Heat Transfer, TU Wien, 1060 Vienna, Austria}
\affiliation{2}{Dipartimento di Ingegneria Meccanica e Aerospaziale, Sapienza Università di Roma, 00184 Rome, Italy}

\correspondingauthor{Marco De Paoli}{\href{mailto:marco.de.paoli@tuwien.ac.at}{marco.de.paoli@tuwien.ac.at}}

\begin{keypoints} 
\item Convective mixing of carbon dioxide in geological formations is time-dependent and universally quantified by mean scalar dissipation
\item Density–concentration relationships and interface mobility strongly modify mixing by altering density contrasts and interface dynamics
\item Common modeling assumptions (fixed interface, 2D flow, monotonic density laws) can significantly bias predictions of CO$_2$ mixing
\end{keypoints}
\justifying
\begin{abstract} 
We investigate how models of fluid properties and boundary conditions influence predictions of convective mixing in confined porous media, with relevance to subsurface carbon dioxide storage. Using high-resolution simulations at high Rayleigh-Darcy numbers (O(10$^4$)), we analyze miscible fluids with linear, nonlinear, and non-monotonic density-concentration relationships under fixed- and free-interface in 2D and 3D. We show that, across all cases, mixing is quantified by the mean scalar dissipation, providing a unifying framework for convective–diffusive interactions. The density-concentration relationship affects mixing via the effective density contrast driving convection and the position of the maximum density. Free interfaces enhance early-time mixing through deformation, while long-term behavior depends on fluid properties and dimensionality. We demonstrate that simplified modeling assumptions (e.g., monotonic density laws or 2D flow) can lead to deviations in predicted mixing rates of up to O(10–100)\%. These results offer guidance for model selection and improving predictions of convective mixing in geophysical systems.
\end{abstract}
\section*{Plain Language Summary} 
When carbon dioxide is injected underground for long-term storage, it can dissolve into salty groundwater and mix through a process driven by small density differences. This mixing process helps trap the carbon dioxide safely, but predicting how fast it occurs is challenging because it depends on how we model the fluids and their interactions. In this study, we use numerical simulations to explore how different assumptions—such as how fluid density depends on concentration, whether the interface between fluids can move, and whether the flow is modeled in two or three dimensions—affect mixing. We find that mixing is controlled by how quickly concentration differences are smoothed out, which is quantified by a quantity called ``scalar dissipation.'' Our results show that some common simplifications can lead to inaccurate predictions. For example, allowing the interface between fluids to move can speed up mixing early on, while realistic three-dimensional flows can behave differently from simpler two-dimensional models. We also identify which fluid properties lead to faster mixing. These findings help improve predictions of carbon dioxide storage in underground formations and provide guidance for designing models of mixing in natural systems.

\section{Introduction}\label{sec:rt_intro}

Convective mixing in porous media plays a central role in a wide range of geophysical processes, including the long-term storage of carbon dioxide (CO$_2$) in subsurface geological formations \cite{schrag2007preparing,szulczewski2013carbon}. 
When CO$_2$ dissolves into brine, small density differences arise, which can trigger buoyancy-driven instabilities, enhancing mixing and accelerating dissolution \cite{huppert2014fluid,Emami-Meybodi2015}. 
This process is widely recognized as a key mechanism controlling the security and efficiency of geological CO$_2$ storage. 
A distinctive feature of CO$_2$–brine systems is the non-monotonic dependence of density on solute concentration, with the density of the mixture exceeding that of both pure fluids, leading to complex and potentially counter-intuitive flow dynamics \cite{hidalgo2015dissolution,boffetta2026free}.

Previous studies have established that mixing in such systems results from the interplay between diffusion and convection, often quantified through the mean scalar dissipation \cite{hidalgo2012scaling}. 
Mixing rates depend sensitively on fluid properties, such as the density-concentration relationship, and on the configuration of the system, including boundary conditions and interface dynamics \cite{hewitt2013convective}. 
In particular, free (mobile) interfaces between CO$_2$ and brine can enhance mixing through deformation and entrainment, while stable density contrasts may suppress interface motion. 
However, most existing studies have considered simplified configurations, typically assuming monotonic density laws \cite{depaoli2025grl}, fixed interfaces \cite{hewitt2013convective,slim2014solutal}, or two-dimensional flow \cite{hidalgo2012scaling,hidalgo2015dissolution}, and their combined impact on mixing predictions remains unclear. 
A key open question is therefore how modeling assumptions (such as the choice of density-concentration relationship, boundary conditions, and flow dimensionality) affect predictions of convective mixing in realistic geophysical settings. 
This issue is particularly relevant for CO$_2$ storage, where fluid composition, chemical interactions, and geological conditions can alter density profiles and flow behavior \cite{thomas2016convective}. 
Understanding the sensitivity of mixing dynamics to these factors is essential for developing reliable predictive models.

In this work, we address this gap by systematically investigating convective mixing in confined porous media using high-resolution numerical simulations. 
We provide a comprehensive assessment of modeling assumptions, namely by investigating both fixed- and free-interface configurations, linear and non-monotonic density-concentration relationships, and two- and three-dimensional flows at high Rayleigh-Darcy numbers.
The resulting database, which we make publicly available \cite{databasethiswork}, provides a valuable benchmark for future theoretical developments, model validation, and comparison with laboratory experiments.
Our study also validates and substantially extends existing theories for predicting mixing by exploring a significantly broader parameter space. Beyond the analysis of individual flow configurations, we identify a unifying framework in which mixing dynamics are governed by the mean dissipation, while fluid properties and interface dynamics control the efficiency and temporal evolution of mixing.
The resulting analysis provides the first systematic assessment of how interface mobility, dimensionality, and density-concentration relationships jointly affect predictions of convective mixing. It identifies the conditions under which commonly adopted modelling simplifications used in the analysis of geological CO$_2$ storage and other subsurface transport processes remain accurate and those requiring more complete descriptions.

\section{Methodology}\label{sec:darcy}

\begin{figure}
    \centering  
    \includegraphics[width=0.99\columnwidth]{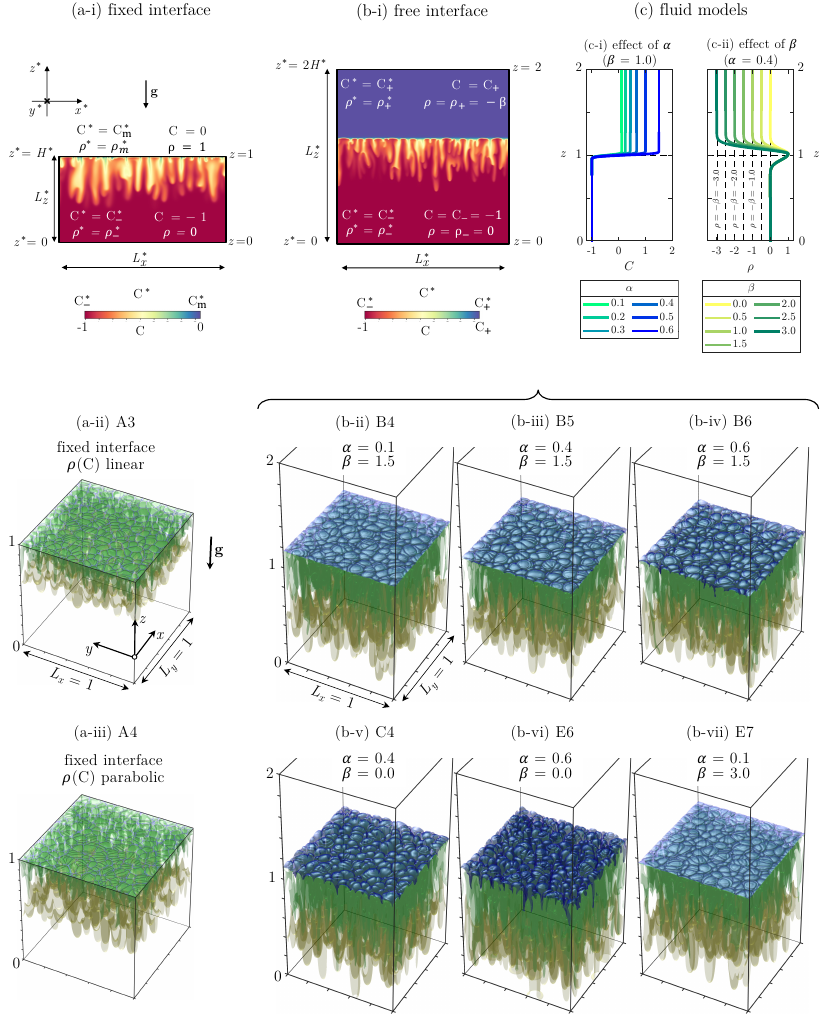}
    \caption{
    Flow configurations considered: (a-i)~fixed interface (discussed in Section~\ref{sec:resA}), and (b-i)~free interface (Section~\ref{sec:resB}). 
    Quantities are shown in dimensional units (superscript $^*$) and dimensionless units (no superscript). 
    Panels~ii-vii: Examples of concentration distribution at time $t=6$; corresponding simulations and parameters are indicated.
    The iso-contours correspond to concentrations $-0.01$ (blue), $-0.6$ (green), and $-0.99$ (yellow).
    The effect of the fluid parameters $\alpha$ and $\beta$ on the fluid concentration and density is illustrated in~(c-i) and~(c-ii), respectively.
    }
    \label{fig:intro0}
\end{figure}

We consider a fluid-saturated porous medium in a two-dimensional domain with uniform porosity $\phi$ and permeability $K$. 
We assume the flow is incompressible and governed by Darcy's law.
A solute, quantified by the concentration $C^*$ (the superscript $^{*}$ denotes dimensional variables), induces variations in the density field $\rho^*$.
The system is periodic in the horizontal directions $x^*,y^*$ and confined by two walls that are impermeable to fluid and solute in the vertical direction $z^*$, along which gravity acceleration $\mathbf{g}$ acts (see Figure~\ref{fig:intro0}a-i,b-i).

At the Darcy scale, the evolution of $C^{*}$, which varies between $C^{*}_-$ and $C^{*}_+$, is governed by the advection–diffusion equation \cite{nield2006convection}
\begin{equation}
\phi\frac{\partial C^{*}}{\partial t^{*}}+\mathbf{u}^{*}\cdot\nabla^{*}C^{*}=\nabla^* \cdot\left(\phi D \nabla^{*} C^{*}\right),
\label{eq:eqadim3}
\end{equation}
where $t^{*}$ is time, $\mathbf{u}^{*}=(u^{*},v^{*},w^{*})$ is the velocity, and $D$ is the molecular diffusion coefficient of the solute. 
We assume that the Boussinesq approximation applies \cite{landman2007heat}.
Then, the flow obeys the continuity and Darcy equations, respectively:
\begin{equation}
\nabla^{*}\cdot\mathbf{u}^{*}=0,
\label{eq:cont}
\end{equation}
\begin{equation}
\mathbf{u}^{*}=-\frac{K}{\mu}\left(\nabla^{*} P^{*}+\rho^{*}g \mathbf{k}\right),
\label{eq:darcy}
\end{equation}
where $\mu$ is the fluid viscosity (assumed constant), $P^{*}$ is the pressure, and $\mathbf{k}$ is the vertical unit vector.

\subsection{Fluids and governing parameters}\label{sec:dens}
We consider $\rho^{*}$ to depend on $C^*$, with a maximum value $\rho^*_m$ at $C^* = C^*_m$ and three $\rho^{*}(C^{*})$ functions: 
\begin{itemize}
\item[(i)] Monotonic linear (Figure~\ref{fig:density}a)
\begin{equation}
\rho^{*}(C^{*})=\rho^{*}_m-(\rho^*_m-\rho^*_-)\frac{C^*_m-C^*}{C^*_m-C_-^*},
\label{eq:dens1}
\end{equation}
with $C_-^*\le C^*\le C^*_m$ and $\rho_-^*\le \rho^*(C^*)\le \rho^*_m$, where $\rho^*_-=\rho^*(C^*_-)$, see \citeA{hewitt2013convective,slim2014solutal,depaoli2025grl}, among others \cite{depaoli2023review}.
\item[(ii)] Monotonic, non-linear (parabolic) (Figure~\ref{fig:density}a)
\begin{equation}
\rho^{*}(C^{*})=\rho^{*}_m-(\rho^*_m-\rho^*_-)\left(\frac{C^*_m-C^*}{C^*_m-C_-^*}\right)^2,
\label{eq:dens2}
\end{equation}
with $C_-^*\le C^*\le C^*_m$ and $\rho_-^*\le \rho^*(C^*)\le \rho^*_m$.
\item[(iii)] Non-monotonic, non-linear, piecewise:
\begin{equation}
\rho^{*}(C^{*})=
\begin{cases}
    \displaystyle\rho^{*}_m-(\rho^*_m-\rho^*_-)\left(\frac{C^{*}_m-C^{*}}{C^{*}_m-C^{*}_-}\right)^2 
    \quad\text{, for } C^*_-\le C^*< C^*_m\\[12pt]
    \displaystyle\rho^{*}_m-(\rho^*_m-\rho^*_+)\left(\frac{C^{*}_m-C^{*}}{C^{*}_m-C^{*}_+}\right)^2 
    \quad\text{, for } C^*_m\le C^*\le C^*_+\\
\end{cases}
\label{eq:dens3}
\end{equation}
with $C_-^*\le C^*\le C^*_+$ and $\rho_+^*\le \rho^*(C^*)\le \rho^*_m$.s
Here $\rho_+^*\le\rho_-^*$; therefore, $\rho_+^*$ represents the lower bound for $\rho^*(C^*)$.
Equation~\eqref{eq:dens3} ensures that at $C^*=C^*_m=0$ the density field is continuous (and maximum, $\rho^*(C^*_m)=\rho^*_m$), with null first derivative, but undefined second derivative.
Different non-monotonic functions have been previously studied \cite{hewitt2013convective,hidalgo2012scaling}.
\end{itemize}

To make quantities dimensionless (indicated by the absence of the superscript $^*$), we choose the convective flow scales, namely $\mathcal{U}^{*}=g (\rho^*_m-\rho^*_-) K / \mu$, $H^*$, and $T^*=\phi H^* /\mathcal{U}^{*}$, while the dimensionless concentration is $C=(C^{*}-C^{*}_m)/(C^{*}_m-C^{*}_-)$.
The governing parameters are:
\begin{equation}
\ra=\frac{\mathcal{U}^{*} H^*}{\phi D}\quad,\quad \alpha=\frac{C^*_+-C^*_m}{C^*_+-C^*_-}\quad,\quad \beta=\frac{\rho^*_- - \rho^*_+}{\rho^*_m-\rho^*_-},
\label{eq:radamain}
\end{equation}
where $\ra$ is the Rayleigh-Darcy number (hereinafter Rayleigh number), and quantifies the relative strength of convective and diffusive mechanisms \cite{hewitt2020vigorous}.
Two additional parameters describing the fluid behavior are $\alpha$ and $\beta$.
The former ($0\le\alpha\le1$) indicates the position of $C_m$, with $C^*_m=C^*_+$ for $\alpha=0$ and $C^*_m=C^*_-$ for $\alpha=1$ (see Figure~\ref{fig:intro0}c-i).
The latter, $\beta$ (see Figure~\ref{fig:intro0}c-ii), quantifies the density contrast between the initial fluid layers ($\rho^*_- - \rho^*_+$) relative to the maximum density contrast ($\rho^*_m-\rho^*_-$).
See also Figures~\ref{fig:density}(c) and \ref{fig:density}(b).

Equations~\eqref{eq:eqadim3}-\eqref{eq:darcy}, with the appropriate $\rho^*(C^*)$ (Equations~\eqref{eq:dens1}, \eqref{eq:dens2}, or \eqref{eq:dens3}), are solved numerically in dimensionless form using a parallel solver \cite{pirozzoli2021towards,depaoli2022strong,depaoli2025grl}.
Additional details on fluid models and solver are provided in Text~S\ref{sec:appA} and Text~S\ref{sec:dimless}, respectively, in Supporting Information S1.

\subsection{Flow configurations}\label{sec:config}
The domain extension is $L^*_x=(L^*_y=)L^*$ in the horizontal directions. 
A sketch of the computational domain, including the boundary conditions (b.c.s), is shown in Figure~\ref{fig:intro0}.
The walls located at $z^*=0$ and $z^* = L_z^*$ are impermeable to the fluid, and no-penetration b.c. apply. Periodicity is enforced in the wall-parallel directions ($x^* = 0$, $x^* = L_x^*$).
We consider two configurations: 
\begin{itemize}
    \item[(i)] Fixed interface: $L_z^*=H^*$, and the concentration is fixed at the top boundary, while a no-flux b.c. is imposed at the bottom.
    The b.c.s read:
\begin{equation}
    \left\{ \begin{array}{ll}
\displaystyle\frac{\partial C^*}{\partial z^*}=0 \quad, \quad\mathbf{u}^* \cdot \mathbf{n} = 0 \qquad \text{at } z^*=0 \\[8pt]
\displaystyle C^*=C^*_m \quad, \quad\mathbf{u}^* \cdot \mathbf{n} = 0 \qquad \text{at } z^*=H^* 
    \end{array}\right.
    \label{eq:bc2a}
\end{equation}
with $\mathbf{n}$ the unit vector normal to the boundary.
    \item[(ii)] Free interface: $L_z^* = 2H^*$, and no-flux b.c. are imposed on both horizontal boundaries:
\begin{equation}
    \left\{ \begin{array}{ll}
\displaystyle\frac{\partial C^*}{\partial z^*}=0 \quad, \quad\mathbf{u}^* \cdot \mathbf{n} = 0 \qquad \text{at } z^*=0 \\[8pt]
\displaystyle \frac{\partial C^*}{\partial z^*}=0 \quad, \quad\mathbf{u}^* \cdot \mathbf{n} = 0 \qquad \text{at } z^*=2H^* 
    \end{array}\right.
    \label{eq:bc2b}
\end{equation}
\end{itemize}

\subsection{Mixing indicators}\label{sec:valid}
Mixing is quantified through the mean scalar dissipation
\cite{hidalgo2012scaling,hidalgo2015dissolution}
\begin{equation}
    \chi = \langle|\nabla C|^2\rangle,
    \label{eq:chi}
\end{equation}
where $\langle\cdot\rangle=1/V\int_V\cdot\,\textrm{d}V$ is the volume average operator.

To quantify and compare the mixing state of systems with different b.c., we propose a modified version of the degree of mixing introduced by \citeA{jha2011quantifying}.
The initial state is characterized by two uniform layers with different concentrations, separated by a sharp interface.
Although there is no upper layer in the fixed-interface configuration, for the sake of comparison with the free-interface, we assume the presence of a virtual layer with the same height as the lower one and constant, uniform concentration.
It follows that the mean concentration variance, defined as $\sigma^2=\langle C^2 \rangle - \langle C \rangle^2$, is initially maximal and equal to $\sigma^2(t=0)=\sigma^2_\text{max}$, and ultimately minimal, corresponding to $\sigma^2(t\to\infty)=0$. 
The variance is representative of the mixing state, which we quantify using the degree of mixing:
\begin{equation}
    M(t) = 1-\frac{\sigma^2(t)}{\sigma^2_\text{max}}.
    \label{eq:degmix}
\end{equation}
This definition implies $0\le M\le 1$, where $M(t=0)=0$ (two segregated layers) and $M(t\to\infty)=1$ (complete mixing).
Details on energy budgets are provided in Text~S\ref{sec:dimless} in Supporting Information S1.

\section{Fixed interface system}\label{sec:resA}
We consider the system in Figure~\ref{fig:intro0}(a-ii), with boundary conditions given by Equation~\eqref{eq:bc2a}. We use this configuration as a reference case to assess the role of dimensionality and fluid properties on mixing, while testing the applicability of existing theoretical models beyond the conditions under which they were originally developed (see \citeA{depaoli2025grl} and references therein).

\subsection{Flow dynamics}\label{sec:resA1b}
The flow dynamics is characterized by three main regimes \cite{slim2013dissolution,slim2014solutal,hewitt2013convective,hidalgo2015dissolution,depaoli2017dissolution,depaoli2025grl}. Initially, a diffusive boundary layer develops at the top of the domain (uniformly at $C=0$, with fixed $C=0$/$\rho=1$ at the top boundary) and thickens until it becomes unstable, forming finger-like structures (Figures~\ref{fig:a3a4}a,b,c,d-i) that transport high-concentration fluid downward.

Once formed, fingers grow at a nearly constant speed (second, convection-dominated regime; Figures~\ref{fig:bulkA}e,f). 2D and 3D dynamics are similar, with slightly faster growth in 3D (Figures~\ref{fig:bulkA}a,b,c,d). The density law $\rho(C)$ affects growth rate: density increases faster for the parabolic case (A4) than the linear one (A3), so fingers reach the bottom sooner in A4 (Figures~\ref{fig:bulkA}b-ii,d-ii,e,f). Growth continues until rising top-boundary concentration reduces the driving density difference - at $t\approx15$ for A3 and $t\approx10$ for A4 - marking the onset of the final, shutdown regime, in which the bulk concentration becomes nearly uniform, driving weakens, and the fingers coarsen (Figures~\ref{fig:bulkA}a,c-iv,v).

\subsection{Analysis of mixing}\label{sec:anmixing1}

\subsubsection{Early-stage dynamics}\label{sec:resA1}
Assuming an initial semi-infinite domain, the concentration profile evolves as:
\begin{equation}
C(z,t)=\text{erf}\left[\frac{(z-1)\ra}{2\sqrt{t\ra}}\right]
\label{eq:diffsol}
\end{equation}
(the equations presented here are derived in Text~\ref{sec:simAflowaa} in Supporting Information S1).
It follows from Equation~\eqref{eq:chi} that
\begin{equation}
    \chi(t) = 
    \sqrt{\frac{\ra}{2\pi t}}.
    \label{eq:mfixchi}
\end{equation}
This diffusive process drives boundary layer thickening and the buildup of a dense fluid layer: both flux and dissipation grow ($0.1<t<0.2$, Figures~\ref{fig:flux1}a,b), and fingers subsequently form. 
For both the linear (A1, A3) and parabolic profiles (A2, A4), the flux growth preceding finger formation occurs earlier in 2D (A1, A2) than in 3D (A3, A4).
As discussed in Section~\ref{sec:resA1b}, for the same concentration field, the density gradients are locally larger for the parabolic profile than for its linear counterpart, leading to larger values of flux and $\chi$ (Figures~\ref{fig:flux1}a,b).

The degree of mixing, defined in Equation~\eqref{eq:mfix2} for the present case, can be computed using Equations~\eqref{eq:diffsol} and~\eqref{eq:mfixchi}:
\begin{equation}
    M(t) =
    \frac{4
    (\sqrt{2}-1)}{\sqrt{\pi\ra}}\sqrt{t} + \frac{4t}{\pi\ra}.
    \label{eq:mfixm}
\end{equation}
In this early-stage phase, $M$ is described by the diffusive solution~\eqref{eq:mfixm} (see inset of Figure~\ref{fig:flux1}c), while the onset of convection closely reflects the dynamics of flux and dissipation, with earlier onset in 2D and for the parabolic profile.

\subsubsection{Convection-dominated and shutdown dynamics}\label{sec:resA2}
The fingers grow at a nearly constant speed \cite{slim2014solutal,depaoli2025grl}, leading to a steady behavior of flux and dissipation (Figure~\ref{fig:flux1}a,b), and subsequently to a linear growth of $M$, which stops at $t\approx t_s$ (shutdown time), due to saturation of the bulk \cite{hewitt2013convective,slim2014solutal}. 
This regime extends over a time span that depends on $\ra$ (the larger $\ra$, the longer this time interval; see~\citeA{slim2014solutal,depaoli2025grl}) and also on the flow dimensionality (3D systems saturate earlier than the corresponding 2D ones, \citeA{depaoli2025grl}).
The finger velocity sets the time required to saturate the bulk, and hence the end of the regime.
This explains the earlier shutdown of convection (occurring at $t_s$ in Figures~\ref{fig:flux1}a,b) in the parabolic profile case (A1, A3, $t_s\approx 10$) compared to the linear ones (A2, A4, $t_s\approx 15$). 

As shown in Figures~\ref{fig:a3a4}(e,f), for $t>t_s$ the bulk is nearly uniform, and the effective concentration difference at the top boundary progressively decreases, considerably slowing down mixing (see Figure~\ref{fig:flux1}c).
Several models have been proposed to describe the dynamics during this regime \cite{slim2013dissolution,slim2014solutal,hewitt2013convective,depaoli2025grl}.
In Text~\ref{sec:shuSI} in Supporting Information S1, we provide a phenomenological description of the underlying physical mechanisms and demonstrate that the model proposed by \citeA{hewitt2013convective}
remains accurate also in three-dimensional flows and for both the linear and parabolic density laws considered here.

\section{Free interface system}\label{sec:resB}
We consider a system defined by the boundary conditions~\eqref{eq:bc2b}, initialized with two layers of different concentration (see Text~S\ref{sec:appA} in Supporting Information S1).
This configuration constitutes the primary focus of the present work, as it provides a more realistic representation of CO$_2$-brine systems and allows us to disentangle the effects of density contrast ($\beta$), shape parameter ($\alpha$), and dimensionality on mixing.

\subsection{Flow dynamics}\label{sec:flowdyn2}
The flow dynamics closely follows the evolution discussed in Section~\ref{sec:resA1b}: an initial diffusive layer at the fluid-fluid interface becomes unstable, forming fingers that merge and transport solute downward, increasing the bulk concentration until the reduced concentration contrast across the interface weakens the driving force. Unlike the fixed-interface case, the mobile, deformable interface here introduces an additional degree of freedom that controls flow evolution and mixing.

We first consider the effect of $\alpha$, which indicates how close the top-layer concentration, $C_+=\alpha/(1-\alpha)$, is to the interfacial concentration ($C_m=0$). Figures~\ref{fig:figbc}(a-c) show simulations B4, B5, and B6 for fixed $\beta=1.5$: smaller $\alpha$ (and hence $C_+$) yields faster flow evolution, which we quantify in Section~\ref{sec:appb_added} via the mean vertical position of the interface. This occurs because smaller $C_+$ produces larger density differences for a given concentration difference, so lower $\alpha$ requires smaller concentration gradients to drive the same density contrast, resulting in faster mixing. In these cases the interface remains weakly deformed and nearly independent of $\alpha$.

\begin{figure} \centering \includegraphics[width=0.99\linewidth]{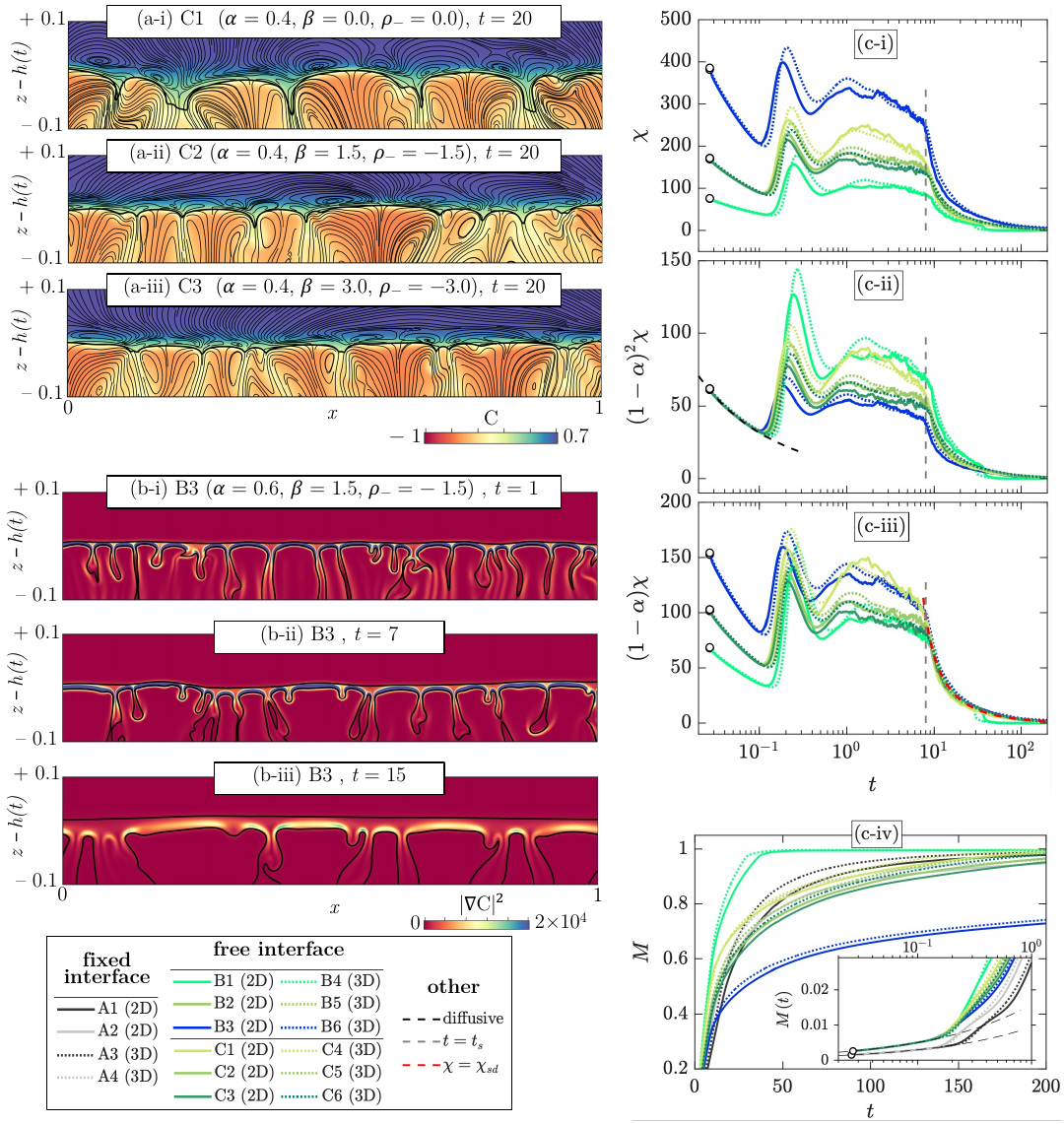} \caption{ 
(a)~Interface topology for $\alpha=0.4$ and different $\beta$ at $t=20$. A portion of domain is shown, centered at the interface height $h$. The instantaneous stream tracers (black thin lines) and the interface (black thick line) are superimposed to $C(x,z)$. The interface is considerably deformed when the initial fluid layers have the same density ($\beta=0$, i.e., a neutrally buoyant case, panel~a-i), making easy for the upwelling (downwelling) plumes to bulge (stretch downwards) the interface. In contrast, for larger $\beta$ (and then a larger density contrast, i.e., a stably stratified case, panels~a-ii and a-iii) the deformation is much lower. Additional details are provided in Fig.~\ref{fig:interface}. 
(b)~Local dissipation $|\nabla C|^2$ for simulation B3 ($\alpha=0.6,\beta=1.5$) at times $t=1$ (b-i), $t=7$ (b-ii) and $t=15$ (b-iii). For better comparison, the maximum concentration is limited to $2\times10^4$, but it can be as high as $2\times10^5$ in (b-i) and (b-ii). Two iso-contours corresponding to $C=1.2$ (upper line) and $C=-0.4$ (lower line) are also shown to visualize the presence of the fingers.
(c)~Mean scalar dissipation $\chi$ and degree of mixing $M$ (solid lines). The shutdown of convection shutdown occurs at $t_s\approx8$ (gray dashed line). Symbols indicated the first time instant considered in the simulations. For ease of visualization, only simulations A-C are shown. $\chi$~(c-i) is rescaled in~(c-ii) to highlight diffusive collapse (compared against Equation~\eqref{eq:mfreechi}, black dashed line). (c-iii)~$\chi$ rescaled to highlight convective shutdown collapse (compared against $\chi_{sd}$ in Equation~\eqref{eq:model3}, red dashed line). B1, B4 differ from the others for $t>30$, when the interface reaches the boundary ($h=2$). (c-iv)~$M$ in the short~(inset) and long~(main panel) term. Simulations A, B and C (solid lines) and the diffusive solutions (Equations~\eqref{eq:mfixm} and~\eqref{eq:mfreem}, for simulations A and B/C, respectively, dashed lines) are shown. } 
\label{fig:3tot} 
\end{figure}

To analyze the effect of the density contrast $\beta$, we fix $C_+=0.7$ ($\alpha=0.4$) and consider simulations C4-C6 (Figures~\ref{fig:figbc}d-f). The interface is strongly affected by $\beta$: larger density differences stabilize the interface, making it harder to deform when $\rho_m-\rho_-$ is large.
Figure~\ref{fig:3tot}(a) illustrates this for three simulations with fixed $\alpha=0.4$ and varying $\beta$. The mean interface height $h(t)$ is defined as the $z$ at which the horizontally averaged concentration $\overline{C}(z)=C_m$. The interface is pulled downward at the finger roots (where $C\approx C_m=0$, Figure~\ref{fig:3tot}a-i), around which the flow is nearly vertical, while rising fluid between fingers pushes the interface upward. For larger $\beta$, the flow above the interface weakens and the interface flattens, with stagnation points visible halfway between consecutive fingers. This dynamics, described by \citeA{hidalgo2015dissolution}, is key to understanding the influence of interfacial dynamics on mixing.

\subsection{Analysis of mixing}\label{sec:anmixing2}
Initially, we have $C(x,z\ge1,t=0)=\alpha/(1-\alpha)$ and $C(x,z<1,t=0)=-1$, which gives $\sigma^2_\text{max}=1/[4(1-\alpha)^2]$.
It follows from Equation~\eqref{eq:degmix} that:
\begin{equation}
    M(t) = \frac{8(1-\alpha)^2}{\ra}\int_0^t\chi\,\text{d}t.
    \label{eq:mfix2b}
\end{equation}
As $t\to\infty$ a uniform concentration field is achieved, and hence $\chi=0$ and $M=1$.

\subsubsection{Early-stage dynamics}\label{sec:resB1}

We consider the initial concentration field (see Text~S\ref{sec:appA} in Supporting Information S1), which varies within the range $0\le C \le C_+=\alpha/(1-\alpha)$.
Assuming a semi-infinite domain in the vertical direction and $\mathbf{u}\approx0$, the solution of Equation~\eqref{eq:equ1bis1} is:
\begin{equation}
C(z,t)=\frac{1}{2(1-\alpha)}\left[2\alpha-1+\text{erf}\left(\frac{(z-1)\ra}{2\sqrt{t\ra}}\right)\right].
\label{eq:diffsol2}
\end{equation}
Using Equation~\eqref{eq:diffsol2} in \eqref{eq:chi}, we obtain an expression for $\chi$:
\begin{equation}
    \chi(t)\approx\frac{1}{(1-\alpha)^2}\frac{1}{4}\sqrt{\frac{\ra}{2\pi t}}.
    \label{eq:mfreechi}
\end{equation}
(a derivation of these equations is provided in Text~\ref{sec:simBCflowaa} in Supporting Information S1).
The evolution of $\chi$ for the free-interface cases is shown in Figure~\ref{fig:3tot}(c-i), and rescaled with the coefficient derived in \eqref{eq:mfreechi} in Figure~\ref{fig:3tot}(c-ii).
The dynamics is well predicted by the analytical solution.
The onset of convection, which is independent of the fluid parameters ($\alpha,\beta$), occurs earlier in 2D than in 3D (Figure~\ref{fig:3tot}c-i), consistently with the fixed-interface configuration.

Using Equations~\eqref{eq:mfreechi} and~\eqref{eq:mfix2b}, the degree of mixing is obtained:
\begin{equation}
    M(t) =
    \sqrt{\frac{8t}{ \pi\ra}},
    \label{eq:mfreem}
\end{equation}
which is independent of the fluid parameters $\alpha$ and $\beta$, as expected in the diffusive phase where density (and hence convection) plays no role.
The early-stage evolution of $M(t)$ for all cases, with fixed interface (simulations A) and with a free interface (simulations B and C), is reported in the inset of Figure~\ref{fig:3tot}(c-iv).
Also in this case, $\alpha$ and $\beta$ do not influence the onset of convection.

\subsubsection{Convection-dominated dynamics}\label{sec:resA3}
The diffusive regime is followed by thickening of the interfacial boundary layer until it becomes unstable, forming fingers. This is first marked by an increase in dissipation departing from diffusive behavior ($0.1 < t < 0.2$), as fingers carry solute away from the interface. Later ($0.2 < t < 0.3$), fingers interact, reorganize, and merge into larger descending structures during a phase of slowly decreasing dissipation, similar to Figure~\ref{fig:flux1}(b). In contrast, dissipation increases with time in the Rayleigh-Taylor-Darcy case \cite{depaoli2025solute,depaoli2019universal}.

Figure~\ref{fig:3tot}(b) shows the local dissipation $|\nabla C|^2$ for simulation B3 at different times. High dissipation is concentrated at the finger boundaries \cite{gopalakrishnan2017relative} and, more dominantly, at the fluid-fluid interface, where \citeA{hidalgo2015dissolution} noted that compression in the interfinger spacing produces large concentration gradients. As the flow evolves, the number of fingers decreases (e.g., from 22 at $t=1$ to 17 at $t=7$, Figures~\ref{fig:3tot}b-i,b-ii), and concentration gradients at the finger sides and across the interface both decrease, explaining the weak decline of $\chi$ in Figure~\ref{fig:3tot}(c-i).

In most cases considered, dissipation in the convection-dominated phase is larger in 3D than in 2D (Figure~\ref{fig:3tot}c-i), consistent with the fixed-interface configuration \cite{fu2013pattern,pau2010high,depaoli2025grl} and attributed to the larger vertical velocities enabled by the additional degree of freedom in 3D. However, \citeA{boffetta2026free} recently reported the opposite trend for $\alpha=0.5$, $\beta=0$, with faster mixing in 2D attributed to greater interface deformation. We confirm this dynamics for simulations C1 and C4 ($\alpha=0.4$, $\beta=0$): the interface is markedly more deformed in 2D than in 3D (see Figure~\ref{fig:interface} and Section~\ref{sec:appb_added}), yielding larger 2D dissipation (Figure~\ref{fig:3tot}c-i). The effect of dimensionality on mixing is nonetheless difficult to characterize due to its transient nature: C1, for instance, exhibits smaller dissipation than C4 prior to the convection-dominated phase.
These results demonstrate that the influence of dimensionality cannot be characterized independently of the fluid properties or of the stage of mixing considered. Rather than being universally more efficient, either two- or three-dimensional flow may promote mixing depending on the combined values of $\alpha$, $\beta$ and the degree of mixing. A quantitative assessment is provided in Section~\ref{sec:concl}.

\subsubsection{Shutdown dynamics}\label{sec:resA4}
When the concentration in the bulk increases, and so does in the boundary layer below the interface, the driving progressively reduces.
As a result, mixing (quantified by $\chi$ in Figure~\ref{fig:3tot}c-i) also decreases. 
The time at which this convective shutdown occurs is $7<t_s<9$, and it also marks a different distribution of low/high dissipation regions in the flow.
Large concentration gradients, indicating large values of $|\nabla C|^2$ and initially localized at the finger boundaries and in the interfinger interfacial regions (Figure~\ref{fig:3tot}b-i,b-ii), are now mostly concentrated at the roots of the fingers (Figure~\ref{fig:3tot}b-iii).

The key role of interfacial dynamics has stimulated the development of physical models capable of predicting its behavior and, simultaneously, describing mixing and bulk concentration.
The model proposed by \citeA{hewitt2013convective}, which we test in detail in Text~\ref{sec:appB} in Supporting Information S1, provides an excellent prediction of the fixed-interface system and also of the free-interface system, but only in the presence of partially miscible fluids and large values of $C_+$, when the interface remains mostly flat and the corresponding boundary layer thickness is uniform.
In our case, in contrast, fluids are fully miscible, and the interface topology evolves following a complex dynamics that this model fails to capture.  

We propose a model to explain the evolution of $\chi$ and $M$ during the shutdown phase.
Dissipation is localized at the interface, with minor contributions from the finger boundaries (Figure~\ref{fig:3tot}b-iii).
Assuming $|\partial_x C|\approx|\partial_y C|\ll|\partial_z C|$, and given an interface thickness $\delta$, $\chi$ can be approximated as:
\begin{equation}
    \chi = \langle |\nabla C|^2 \rangle \approx \frac{\delta}{L_z}\left(\frac{C_+-C_-}{\delta}\right)^2,
    \label{eq:model1}
\end{equation}
with $C_+-C_-=1/(1-\alpha)$.
We assume that $\delta(t)$ depends on $C_+-C_-$ and on a function of time, $f(t)$, that is independent of the fluid parameters: 
\begin{equation}
    \delta(t) = (C_+-C_-)\,f(t)=\frac{f(t)}{1-\alpha}.
    \label{eq:model2}
\end{equation}
The independence of $\delta$ from the parameter $\beta$ is motivated by the fact that interface growth is primarily related to diffusive processes and therefore, to leading order, not directly controlled by the density difference ($\beta$).
Using \eqref{eq:model2} in \eqref{eq:model1}, we obtain that the mean dissipation during the shutdown regime is
\begin{equation}
    \chi_{sd}(t) = \frac{1}{2(1-\alpha) f(t)}, 
    \label{eq:model3}
\end{equation}
where $f(t)=[a_1 + (t-t_s)]/a_2$, with $a_1=3.376$ and $a_2=652.9$ determined from fitting of the numerical results.
The functional form employed for $f(t)$ is obtained from models based on interface deformation proposed by \citeA{hidalgo2015dissolution}.
According to this model, $a_1$ and $a_2$ depend on the effective finger width, $\delta$, $h(t=0)$, $L_x$, $L_y$, and the exponent $n$ of the power law $\rho(C)$ ($n\approx2$).
In the present study, all these parameters, except the effective finger and interface widths, are constant, and therefore $a_1,a_2$ can be interpreted as a measure of those quantities.
Equation~\eqref{eq:model3} provides an excellent estimate of the evolution of dissipation, see Figure~\ref{fig:3tot}(c-iii).
All simulations, with the exception of B1 and B4, which have an interface quickly reaching the upper boundary, follow the predicted $\chi_{sd}$, confirming the accuracy of the proposed model.

The degree of mixing in this phase, reported in the main panel of Figure~\ref{fig:3tot}(c-iv), suggests that higher mixing is achieved for small $\alpha$, e.g., for simulations B1 and B4 ($\alpha=0.1$).
This result, counter-intuitive given the smaller values of dissipation recorded for small $\alpha$ (see Figure~\ref{fig:3tot}c-i), is due to the definition of degree of mixing provided in Equation~\eqref{eq:mfix2b}, which accounts not only for $\chi(t)$ but also for $\alpha$. 
The evolution of $M$ describes, through a single global indicator, the evolution of the system and will be used to compare different configurations.

\section{Discussion and conclusions}\label{sec:concl}
With respect to geological CO$_2$ sequestration, the modeling framework considered is idealized in several aspects. In particular, the fluids miscibility represents a strong assumption. This, as well as other potential limitations, are discussed in detail in Section~\ref{sec:limout}.
Within the fully miscible scenario considered, the results presented above show that the impact of commonly adopted modelling assumptions depends strongly on the stage of the mixing process. Since the shutdown regime accounts for 50–80\% of the overall mixing, understanding its dynamics is essential for predicting long-term convective dissolution.
This is quantified via the degree of mixing in the main panel of Figure~\ref{fig:3tot}(c-iv), which depends on the history of the system.
Therefore, it is difficult to interpret the late-time flow dynamics without accounting for the previous phases contributing to mixing. 
To analyze and compare the role of different effects, we consider the time $t_\%$ required to achieve a prescribed degree of mixing $M_\%$. For instance, $t_{30\%}$ corresponds to the time required to reach $M=0.3$.

\begin{figure}
    \centering
    \includegraphics[width=0.99\linewidth]{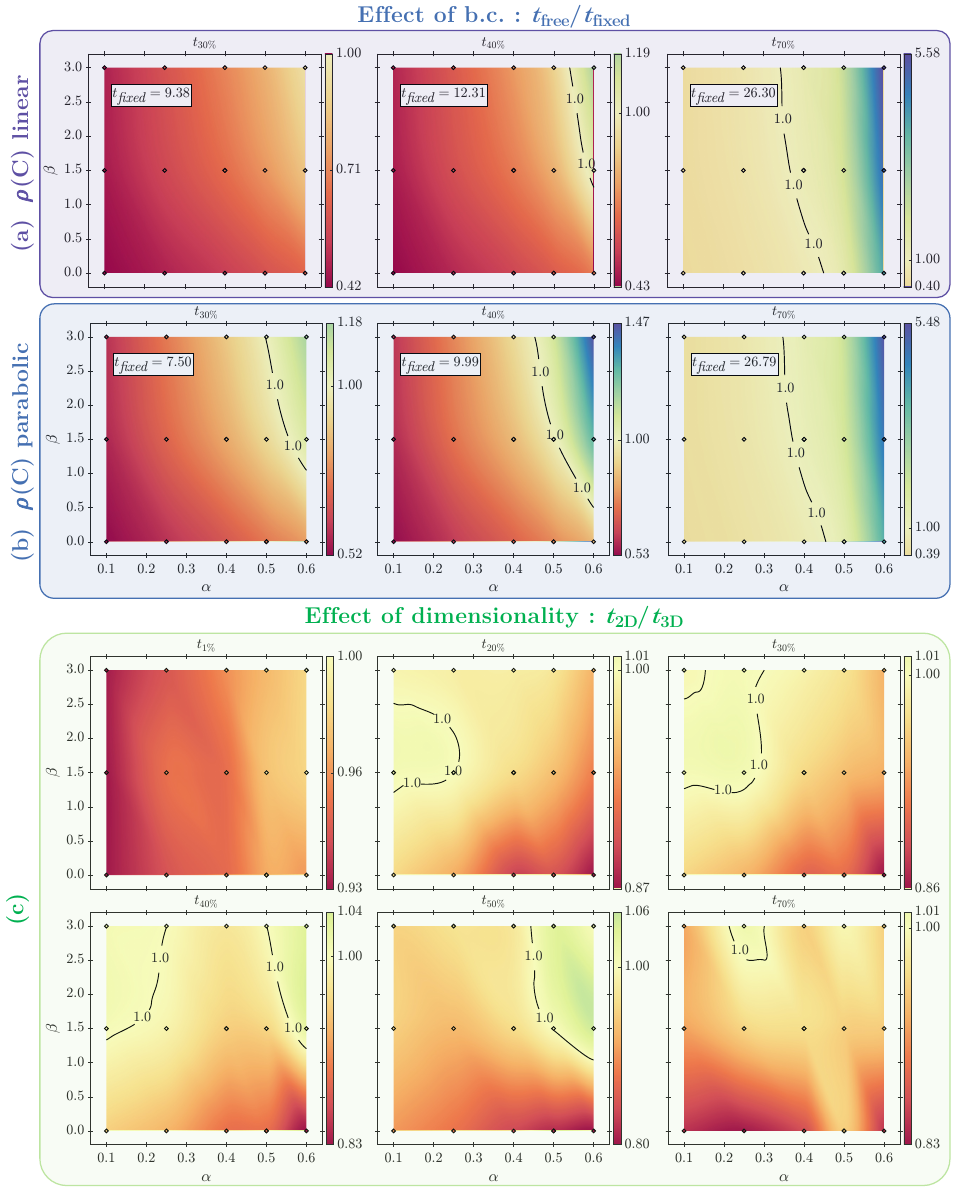}       
    \caption{
    (a,b)~Comparison of times required to achieve a given degree of mixing in 3D with the free interface (as a function of $\alpha,\beta$) relative to the fixed-interface case (e.g., $t_{30\%}$ indicates the time required to achieve $M=0.3$).
    The value of time corresponding to $t_\%$ obtained in the fixed-interface system, $t_\text{fixed}$, is indicated in each panel and refers to (a)~linear $\rho(C)$ and (b)~parabolic $\rho(C)$.
    (c)~Comparison of times required to achieve a prescribed degree of mixing in 2D relative to 3D in the free-interface case.
    The iso-contour corresponding to a time ratio equal to 1 is indicated with a thick black line.
    }
    \label{fig:time_tot_GRL}
\end{figure}

The results reported in Figure~\ref{fig:time_tot_GRL} highlight the role of modeling assumptions (monotonic $\rho(C)$, 2D, fixed interface) relative to the most complete model considered here (3D, non-monotonic $\rho(C)$, free interface). The following effects emerge:
\begin{itemize}
    \item[i)] \textbf{Interface model:} 
    Despite the dependence of $M(t)$ on $\alpha$ and $\beta$, mixing is initially more efficient in free-interface systems (Figure~\ref{fig:3tot}c-iv).
    However, at long times this is, in general, no longer true and depends on the flow parameters.
    3D results in Figure~\ref{fig:time_tot_GRL}(a,b) confirm this, indicating that $t_{30\%}$ is smaller for the free-interface case compared to the fixed-interface case. For higher degrees of mixing ($t_{50\%}, t_{70\%}$), the threshold value for which the free- and fixed-interface systems exhibit the same $t_\%$ depends on the fluid parameters $(\alpha,\beta)$, with $\alpha$ playing a leading role.
    \item[ii)] \textbf{Dimensionality:}
    The onset of convection occurs earlier in 2D than in 3D (see Figure~\ref{fig:3tot}c-i), and this effect is clearly reflected in the time ratio at $t_{1\%}$ reported in Figure~\ref{fig:time_tot_GRL}(c). 
    The effects of the fluid properties on the interface deformation (and then on mixing) become more significant after the fingers have formed and grown. In 2D, fingers are coherent sheet-like concentration regions constrained to a plane. In 3D, fingers can deform, twist, split, and interact also in the third direction. This extra degree of freedom promotes fragmentation into smaller fingers and a more complex flow network, which impairs coherence of the fingers \cite{van2013comparison}, thus their strength and ability of deforming the interface. However, interface deformation also changes across the parameters space (see Fig.~\ref{fig:interface} and Movie~S4): it is more pronounced when the variation of the concentration across the interface is gradual (large $\alpha$) or the density contrast between the two layers is small (small $\beta$). 
    Therefore, the dissipation during the convection-dominated phase may be higher in 3D than in 2D, or vice versa (e.g., for $\alpha=0.5$ and $\beta=0$, in agreement with \citeA{boffetta2026free}), depending on the fluid parameters $(\alpha,\beta)$ (see Figure~\ref{fig:3tot}c-ii). 
    The cumulative mixing in these regimes yields a complex, time-dependent picture: for all times considered ($t_{20\%}$–$t_{70\%}$, Figure~\ref{fig:time_tot_GRL}c), two distinct regions of the parameter space $(\alpha,\beta)$ exist in which mixing, depending on the fluid properties, three-dimensional flow may accelerate mixing by up to 20\%, whereas in other regions of parameter space it may instead delay it by up to 6\%.  
    \item[iii)] \textbf{Fluid properties:} 
    The most effective parameter combination for faster mixing can be derived from Figures~\ref{fig:time_tot_GRL}(a,b).
    The dominant parameter is $\alpha$: the smaller $\alpha$, the sooner a given degree of mixing is achieved, in agreement with previous observations by \citeA{hidalgo2012scaling,soboleva_simulation_2025}.
    However, for intermediate times (e.g., $t_{40\%}$), the role of $\beta$ is also important.
    Finally, we observe in Figures~\ref{fig:time_tot_GRL}(a,b) that the density law considered in the fixed-interface case may substantially influence mixing at intermediate stages ($t_{30\%},t_{40\%}$), while no major difference is observed for larger degrees of mixing ($t_{70\%}$).
\end{itemize}
One simple outcome of this work is that optimal conditions for mixing are achieved for small $\alpha,\beta$ (see Figures~\ref{fig:time_tot_GRL}(a,b)), for which the interface is highly deformed.

Beyond the specific findings reported here, this work establishes a unified quantitative framework for the analysis of convective mixing in porous media. Rather than considering individual physical mechanisms or modeling choices separately, it systematically combines interface mobility, dimensionality, and density-concentration relationships within a common numerical framework, enabling their relative influence on mixing dynamics to be assessed over a substantially broader parameter space than previously explored. This unified perspective not only extends and consolidates existing theories but also provides practical guidelines for selecting appropriate modeling assumptions and identifying the conditions under which simplified descriptions remain accurate or more complete formulations are required. Together with the openly available database \cite{databasethiswork}, it establishes a valuable benchmark for future theoretical developments, model validation, and the interpretation of experiments involving miscible or partially miscible fluids in porous media \cite{backhaus2011convective,neufeld2010convective,croccolo2024,imuetinyan2024transparent,xu2026comparison}.

\subsection{Limitations and future developments}\label{sec:limout}
When compared to convective mixing in geological CO$_2$ sequestration sites, the conditions considered in the present study are idealized in several aspects, including the assumptions of homogeneity \cite{chen2015diffuse,hidalgo2022advective,hu2024double}, absence of fractures \cite{keim2025rayleigh,ulloa2025convection}, dispersion \cite{hidalgo2009effect,depaoli2025solute,depaoli2024towards,liang2018,wen2018rayleigh}, and chemical reactions \cite{dewit2004miscible,jotkar2019enhanced,dewit2020chemo,strauch2025measuring}. 
Moreover, fluids are considered fully miscible, whereas CO$_2$ and brine are only partially miscible.
As a result, the large density gradient across the resulting CO$_2$–brine interface (corresponding to a large $\beta$ in the present model) remains sharp, inducing a nearly flat interface behavior (see, for instance, simulation~E7 in Figure~\ref{fig:intro0}b-vii). 
Partially miscible fluids have been modeled in 2D by \citeA{li2022diffuse,li2023dissolution} using a phase-field method (Darcy–Cahn–Hilliard).
When comparing miscible and partially miscible cases, they observed that in the convective phase mixing is higher in the miscible case, and convective shutdown occurs earlier.
Their results suggest that any extension of findings from miscible to partially miscible systems should be treated with caution.
A possible development of this work therefore consists in analyzing partially miscible systems and investigating the role of dimensional effects on flow and mixing evolution.

\section*{Conflict of Interest}
The authors declare no conflicts of interest relevant to this study.
\section*{Open Research Section}
The data of dissipation and interface elevation presented in this work are available at \citeA{databasethiswork}.

\acknowledgments
Funded by the European Union (ERC, MORPHOS, 101163625). Views and opinions expressed are however those of the author(s) only and do not necessarily reflect those of the European Union or the European Research Council. Neither the European Union nor the granting authority can be held responsible for them.
The authors acknowledge the TU Wien Bibliothek for financial support through its Open Access Funding Programme.
The computational results presented have been achieved using in part the Vienna Scientific Cluster (VSC).

\bibliography{bibliography}
\section*{References From the Supporting Information}
\renewcommand{\refname}{\vspace{-\baselineskip}}


\end{document}